\input{aipcheck}

\documentclass[
    ,final            
  ]
  {aipproc}

\layoutstyle{6x9}

\begin{document}

\title{Excited $L=1$ baryons in large $N_c$ QCD}

\author{Dan Pirjol}{
  address={Dept. of Physics and Astronomy, The Johns Hopkins University,
3400 N. Charles Street, Baltimore, MD 21218}
}

\author{Carlos Schat}{
  address={Department of Physics, Duke University, Durham,
NC 27708}
}


\begin{abstract}
The physics of the orbitally excited baryons simplifies
drastically in the large $N_c$ limit. The states are arranged
into irreducible representations of the contracted $SU(4)_c$ 
symmetry, with mixing angles determined exactly. The ratios of the strong 
couplings $N^{*}\to [N\pi]_{S,D}$ are predicted in this limit, 
with results in agreement with those following from the quark 
model (with the large $N_c$ mixing angles). 
We present a phenomenological
analysis of the observed nonstrange baryons from the perspective
of the $1/N_c$ expansion, including constraints from their masses
and strong decays. 
\end{abstract}

\maketitle



It has been known for some time that the large $N_c$ limit of QCD 
\cite{1,2} can give a useful qualitative description of low energy
hadronic physics.
In the baryon sector this limit turns out to be considerably more
predictive \cite{DM1,DM2,J}, and the $1/N_c$ expansion 
can be formulated in a systematic way allowing the treatment of power
corrections and SU(3) breaking effects \cite{DJM1,LuMa,DJM2} (for a
recent review see \cite{Jreview}).

The crucial point is the emergence of a new symmetry of
QCD in the large $N_c$ limit of the baryon sector - the
contracted $SU(2n_f)_c$ symmetry (with $n_f$ the number of light flavors)
\cite{DJM1}. The physical states arrange themselves in irreducible 
representations of this symmetry group, which for $n_f=2$ are labeled 
by $K=0,\frac12,1,\dots$ and contain all states satisfying $|I-J| \leq K$. 
The leading order predictions for masses and strong couplings recover the 
quark model with $SU(2n_f)$ spin-flavor symmetry. 
Using this approach, many applications have been discussed for the ground 
state baryons \cite{Jreview,JJM,quad}.

The large $N_c$ expansion has been applied also to excited 
baryons, using different implementations of the idea
\cite{CGKM,JG,PY1,PY2,CCGL1,CCGL,CC1,CC2,SGS,su3prd,su3prd1}. 
In Refs.~\cite{PY1,PY2} the contracted $SU(4)_c$ symmetry was found to 
extend also to these states, using consistency conditions for $N^{(*)}\pi 
\to N^{(*)}\pi$ scattering.
However, because of the more complex mass spectrum of the excited states,
the implications of this symmetry are more rich than in the ground
state sector. In particular, the nonstrange negative parity $L=1$ baryons 
fall into three irreducible representations of the $SU(4)_c$ symmetry
\begin{eqnarray}\label{towers}
K=0: & & N_{1/2}, \Delta_{3/2}, \dots\\
K=1: & & N_{1/2}, N_{3/2}, \Delta_{1/2}, \Delta_{3/2}, \Delta_{5/2}, 
\dots\nonumber\\
K=2: & & N_{3/2}, N_{5/2}, \Delta_{1/2}, \Delta_{3/2}, \Delta_{5/2},
\Delta_{7/2}, \dots\,.\nonumber
\end{eqnarray}
This can be contrasted with the case of the corresponding
ground state baryons, which include only one representation $K=0:\,
N_{1/2}, \Delta_{3/2},\dots$. Thus, the large $N_c$ limit implies a
mass pattern for the excited baryons Eq.~(\ref{towers}) 
 which is very 
different from the quark model prediction of complete degeneracy into
the {\bf 70} of $SU(6)$ \cite{su6,mix}. Still, the large $N_c$ predictions
for $N^*\to N\pi$ amplitude ratios are found to be again in agreement 
with those of
the quark model with SU(4) spin-flavor symmetry \cite{PY1}. [We note that 
very similar predictions are
obtained for hybrid baryons in the large $N_c$ limit \cite{CPY}.]

In a recent paper \cite{PS}, the status of the $1/N_c$ expansion for 
the nonstrange $L=1$ baryons was reexamined, working consistently at 
leading and subleading order in $1/N_c$. This was done using the operator
approach proposed in \cite{LuMa,DJM1}, and first applied to the excited
states in \cite{JG,CCGL1,CCGL}.
The mass matrix of the $L=1$ baryons can be written as a sum of operators
acting on the quark basis as
\begin{eqnarray}\label{massop}
\hat M = \sum_{k=0}^{N_c} \frac{1}{N_c^{k-1}} C_k {\cal O}_k
\end{eqnarray}
with ${\cal O}_k$ a $k$-body operator. Both the coefficients $C_k$ 
and the matrix elements of the operators on baryon states $\langle {\cal O}_k\rangle$
have power expansions in $1/N_c$ with coefficients determined by nonperturbative
dynamics
\begin{eqnarray}
C_k = \sum_{n=0}^\infty \frac{1}{N_c^n} C_k^{(n)}\,, \qquad
\langle {\cal O}_k \rangle = \sum_{n=0}^\infty
\frac{1}{N_c^n}\langle {\cal O}_k 
\rangle^{(n)} \,.
\end{eqnarray}
The natural size for the coefficients $C_k^{(n)}$ is $\Lambda \sim 500 $ MeV.
A complete basis for the operators ${\cal O}_k^{(1,2)}$ has been constructed in 
\cite{CCGL}, to which we refer for further details.
At leading order in $N_c$ only three operators contribute to the mass matrix,
given by
\begin{eqnarray}
O_1 = N_c {\bf 1}\,,\qquad
O_2 = l^i s^i\,,\qquad
O_3 = \frac{3}{N_c}l^{(2)ij} g^{ia} G_c^{ja}\,.
\end{eqnarray}
At subleading order   $O(N_c^{-1})$ five additional operators
start contributing
\begin{eqnarray}
O_4 = l s + \frac{4}{N_c+1} l t G_c\,,\,\,
O_5 = \frac{1}{N_c} l S_c\,,\,\, 
O_6 = \frac{1}{N_c} S_c S_c\,,\,\,
O_7 = \frac{1}{N_c} s S_c\,,\,\,
O_8 = \frac{1}{N_c} l^{(2)} s S_c\,.
\end{eqnarray}
Their matrix elements on the excited baryon states can be found in the
Appendix of \cite{CCGL}.
These operators have a direct physical interpretation in the quark model
in terms of one- and two-body quark-quark  couplings. 

Keeping only the operators $O_{1,2,3}$ contributing 
at $O(N_c^0)$, one finds by direct diagonalization of the mass matrix
the mass eigenstates in the large $N_c$ limit as linear combinations of
the quark model $N_{1/2}, N'_{1/2}$ states 
\begin{eqnarray}\label{J1/2}
\begin{array}{l}
|K = 0\,, J=\frac12 \rangle = 
\frac{1}{\sqrt3} N_{1/2} + \sqrt{\frac23} N'_{1/2} \\
|K = 1\,, J=\frac12 \rangle =
-\sqrt{\frac23} N_{1/2} + \frac{1}{\sqrt3} N'_{1/2} \\
\end{array}
\qquad
\left\{
\begin{array}{l}
M_0^{(0)} = N_c C_1^{(0)} - C_2^{(0)} - \frac{5}{8} C_3^{(0)} \\
M_1^{(0)} = N_c C_1^{(0)} - \frac12 C_2^{(0)} + \frac{5}{16} C_3^{(0)} \\
\end{array}
\right.
\end{eqnarray}
A similar diagonalization of the mass matrix for the $J=\frac32$ $N^*$ states 
gives the eigenstates
\begin{eqnarray}\label{J3/2}
\begin{array}{l}
|K = 1\,, J=\frac32 \rangle =
\frac{1}{\sqrt6} N_{3/2} + \sqrt{\frac56} N'_{3/2} \\
|K = 2\,, J=\frac32 \rangle = 
-\sqrt{\frac56} N_{3/2} + \frac{1}{\sqrt6} N'_{3/2} \\
\end{array}
\qquad
\left\{
\begin{array}{c}
M_1^{(0)} = N_c C_1^{(0)} - \frac12 C_2^{(0)} + \frac{5}{16} C_3^{(0)}\\
M_2^{(0)} = N_c C_1^{(0)} + \frac12 C_2^{(0)} - \frac{1}{16} C_3^{(0)}\\
\end{array}
\right.
\end{eqnarray}
The $N_{5/2}$ state does not mix and has the mass $M_2^{(0)}$.
These results make the tower structure in Eq.~(\ref{towers}) explicit.

There is a discrete ambiguity in the assignment of the
five observed $N^*$ excited nucleons into the large $N_c$ irreducible reps
of $SU(4)_c$. The four possible ways of grouping them into multiplets are 
shown in Table \ref{assign}.
This implies a four-fold ambiguity in the
coefficients of the mass operator $C_i^{(0)}$. In the following we 
extract these coefficients and attempt to resolve the discrete ambiguity by using 
experimental information on masses and strong decays of these states.

\begin{table}
\begin{tabular}{c|c|c|c|c}
\hline
 & K = 0 & K = 1 & K = 2 & ordering\\
\hline
\hline
\mbox{\# 1} & $N_{1/2}(1650)$ & $\{N_{1/2}(1535), N_{3/2}(1520)\}$ &  
$\{N_{3/2}(1700), N_{5/2}(1675)\}$ & $\{M_0\,, M_2\} > M_1$ \\
\mbox{\# 2} & $N_{1/2}(1535)$ & $\{N_{1/2}(1650), N_{3/2}(1520)\}$ &  
$\{N_{3/2}(1700), N_{5/2}(1675)\}$ & $M_2 > M_1 > M_0$\\
\mbox{\# 3} & $N_{1/2}(1535)$ & $\{N_{1/2}(1650), N_{3/2}(1700)\}$ &  
$\{N_{3/2}(1520), N_{5/2}(1675)\}$ & $M_1 > \{M_0\,, M_2\}$\\
\mbox{\# 4} & $N_{1/2}(1650)$ & $\{N_{1/2}(1535), N_{3/2}(1700)\}$ &  
$\{N_{3/2}(1520), N_{5/2}(1675)\}$ & $ M_0 > M_1 > M_2$\\
\hline
\end{tabular}
\caption{The four possible assignments of the observed nonstrange excited
baryons into large $N_c$ towers with $K = 0, 1, 2$.}
\label{assign}
\end{table}

We start by determining the values of the coefficients $C_{1,2,3}^{(0)}$ 
in the large $N_c$ limit, using the mass eigenvalues given in 
Eqs.~(\ref{J1/2}) and (\ref{J3/2}).  For each assignment, we fitted the 
coefficients $C_{1,2,3}^{(0)}$ to the observed $N^*$ masses \cite{PS}. 
The results for $C_{2,3}^{(0)}$ are shown graphically in
Figure 1 for each of the four possible assignments. The mixing angles 
$\theta_{N1}, \theta_{N3}$ are fixed by Eqs.~(\ref{J1/2}) and (\ref{J3/2}).

\begin{figure}\label{fig}
  \includegraphics[height=.3\textheight]{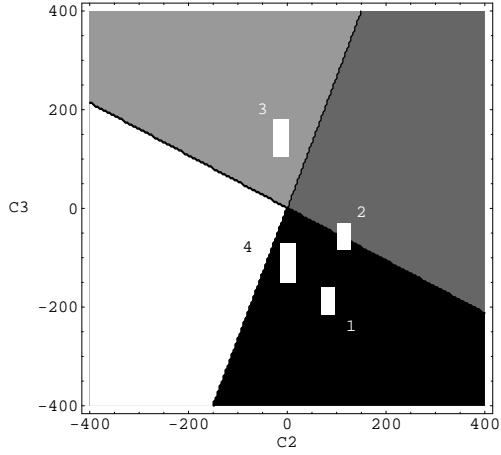}
  \caption{Fit results for the leading order coefficients $C_{2,3}^{(0)}$,
corresponding to each assignment. The four wedgelike regions show the allowed 
values for each assignment following from the noncrossing argument.}
\end{figure}

These results must satisfy an additional constraint, following
from the no-crossing property of the eigenstates with the same quantum
numbers. Consider the masses of the two $J=1/2$ states as functions of
$1/N_c$. They can not cross when $N_c$ is taken from 3 to infinity. This means
that the correspondence of the physical $N_{1/2}$ states with the large $N_c$
towers is fixed by the relative ordering of the $K=0,1$ towers. 
This leads to a connection between the ordering of the tower masses and each
of the 4 assignments, shown in the last column of Table \ref{assign}.
This constraint is also shown graphically in Fig.~\ref{fig}; it rules out the
assignment No.4 and further restricts the solution for the assignment No.2.

We consider next also information from the strong decays $N^* \to [N\pi]_{S,D}$.
The large $N_c$ predictions for these decays were given in \cite{PY1},
where the consistency conditions for $N^{(*)}\pi \to N^{(*)}\pi$
scattering were solved exactly. Referring to Ref.~\cite{PY1,PY2} for the full 
solution, we list in  Eqs.~(\ref{SDwave}) the results
for the S- and D-wave reduced amplitudes $A_{\rm red}$ [defined up to
spin and isospin CG coefficients as $A(N^*\to N\pi) = 
A_{\rm red} \cdot CG_I \cdot CG_J$].
The $0$'s in these relations denote $1/N_c$ suppressed amplitudes.
Including spin and isospin factors, these relations predict
the large $N_c$ partial width ratios shown in Eq.~(\ref{SDwidths}).
In addition to constraining the masses of the tower states, the
contracted $SU(4)_c$ symmetry relates also their strong decay widths
which are predicted to be equal. This equality holds also for
individual channels, which implies sum rules such as (for the $K=2$ states)
\begin{equation}
\Gamma(N_{3/2}\to [N\pi]_D) + \Gamma(N_{3/2}\to [\Delta\pi]_D) =
\Gamma(N_{5/2}\to [N\pi]_D) + \Gamma(N_{5/2}\to [\Delta\pi]_D)\,.
\end{equation}
These relations are broken by $1/N_c$ terms in the expansion of the
$N^*\to N$ axial current, and by kinematical phase space effects.

\begin{eqnarray}\label{SDwave}
\begin{array}{lll}
K= 0 & K = 1  & K=2 \\
\hline
(N_{\frac12} \to [N \pi]_S) = 0 &
(N_{\frac12} \to [N \pi]_S) = \sqrt2 c_S & 
(N_{\frac32} \to [\Delta \pi]_{S}) = 0 \\
 & (N_{\frac32} \to [\Delta \pi]_{S}) =  c_S & \\
\hline
(N_{\frac12} \to [\Delta \pi]_D) = 0 &
(N_{\frac12} \to [\Delta \pi]_D) = c_{D1} &
 (N_{\frac32} \to [N \pi]_D) = c_{D2}\\
 & (N_{\frac32} \to [N \pi]_{D}) = - 2 c_{D1} &
(N_{\frac32} \to [\Delta \pi]_{D}) = -\frac12 c_{D2} \\
 & (N_{\frac32} \to [\Delta \pi]_{D}) =  - c_{D1} &
(N_{\frac52} \to [N \pi]_{D}) =  \sqrt{\frac23} c_{D2} \\
 & &
(N_{\frac52} \to [\Delta \pi]_{D}) =  \frac12\sqrt{\frac73} c_{D2} \\
\hline
\end{array}
\end{eqnarray}

Note that these predictions depend crucially on the $K$ assignment
of the excited baryons. In particular, the strong couplings of the $K=0$
states are suppressed by $1/N_c$. Also, the $J=3/2$
$K=2$ state is predicted to decay in a pure $D-$wave. Therefore one expects 
these predictions to be useful for distinguishing
among  the possible assignments.

\begin{eqnarray}\label{SDwidths}
& & K=1:\, \Gamma(N_{\frac12} \to [N \pi]_S) : 
\Gamma(N_{\frac32} \to [\Delta \pi]_S) = 1 : 1\\
& & K = 1:\, \Gamma(N_{\frac12} \to [\Delta \pi]_D) :
\Gamma(N_{\frac32} \to [N \pi]_D) :
\Gamma(N_{\frac32} \to [\Delta \pi]_D)  = 2 : 1 : 1\nonumber\\
& & K = 2:\, \Gamma(N_{\frac32} \to [N \pi]_D) :
\Gamma(N_{\frac32} \to [\Delta \pi]_D) :
\Gamma(N_{\frac52} \to [N \pi]_D) : \Gamma(N_{\frac52} \to [\Delta \pi]_D)\nonumber\\
& & \hspace{4cm}  = \frac12 : \frac12 : \frac29 : \frac79  \,.\nonumber
\end{eqnarray}

For this purpose, we consider the ratios of $S-$wave partial widths
$R_1 = \frac{\Gamma(N_{1/2}(1535)\to N\pi)}{\Gamma(N_{3/2}(1520)\to [\Delta\pi]_S)}$ 
and $R_2 = \frac{\Gamma(N_{1/2}(1650)\to N\pi)}{\Gamma(N_{1/2}(1535)\to N\pi)}$. 
We present in Table \ref{strong} the large $N_c$ predicted values
for $R_{1,2}$ (including phase space factors), together with their experimental values.

\begin{table}
\begin{tabular}{c|ccc}
\hline
 & $R_1$ & $R_2$ & $R_1 R_2$ \\
\hline
\hline
\mbox{\# 1} & 2.05 & $O(1/N_c^2)$ & $O(1/N_c^2)$ \\
\mbox{\# 2} & $O(1/N_c^2)$ & $O(N_c^2)$ & 2.4  \\
\mbox{\# 3} & $O(1)$ & $O(N_c^2)$ & $O(N_c^2)$ \\
\mbox{\# 4} & $O(N_c^2)$ & $O(1/N_c^2)$ & $O(1)$ \\
\hline
exp & 3.6-13.7 & 1.0-2.6 & 5.7-22.5 \\
\hline
\end{tabular}
\caption{Large $N_c$ predictions for the ratios of strong decay widths
$R_{1,2}$, and their experimental values.}
\label{strong}
\end{table}

Despite the large experimental errors, the combined constraints from
masses and strong decays ($R_{1,2}$) appear to favor the assignment No.1 
\cite{PY1,PS}. 
In Ref.~\cite{PS} the mass analysis presented here was extended to $O(1/N_c)$.
The most important new point of this analysis is the appearance of a continuous
set of solutions for the mass operator. Including also data from excited 
$\Delta$ states, it was found that the assignment No.3 is favored, in agreement 
with the analysis in \cite{CCGL}, although No. 1 is still marginally allowed. 
More conclusive results will be possible once better data on the masses and
decay widths of these states will become available.

Finally, we note that similar conclusions on the mass spectrum of these states
(\ref{towers}) were also reached in \cite{CoLe1,CoLe2,CoLe3} from a study of
$N\pi$ scattering amplitudes in the Skyrme model.


\begin{theacknowledgments}

This work was supported by the DOE and NSF under Grants No. 
DOE-FG03-97ER40546 (D.P.),
NSF PHY-9733343, DOE DE-AC05-84ER40150 and DOE-FG02-96ER40945 (C.S.).
\end{theacknowledgments}


\bibliographystyle{aipproc}   

\bibliography{sample}

\IfFileExists{\jobname.bbl}{}
 {\typeout{}
  \typeout{******************************************}
  \typeout{** Please run "bibtex \jobname" to optain}
  \typeout{** the bibliography and then re-run LaTeX}
  \typeout{** twice to fix the references!}
  \typeout{******************************************}
  \typeout{}
 }

\end{document}